# Micro-solvation of CO in water: Infrared spectra and structural calculations for $(D_2O)_2$ - CO and $(D_2O)_3$ - CO

*Aaron J. Barclay,[1] Andrea Pietropolli Charmet,[2] Kirk H. Michaelian,[3] A.R.W. McKellar,[4] and Nasser Moazzen-Ahmadi[1,*]*

[1]Department of Physics and Astronomy, University of Calgary, 2500 University Drive North West, Calgary, Alberta T2N 1N4, Canada

[2]Dipartimento di Scienze Molecolari e Nanosistemi, Università Ca' Foscari Venezia, Via Torino 155, I-30172, Mestre, Venezia, Italy

[3]CanmetENERGY, Natural Resources 1 Oil Patch Drive, Suite A202, Devon, Alberta T9G 1A8, Canada

[4]National Research Council of Canada, Ottawa, Ontario K1A 0R6, Canada.

AUTHOR INFORMATION

**Corresponding Author**

*N. Moazzen-Ahmadi.

nmoazzen@ucalgary.ca.

Department of Physics and Astronomy, University of Calgary, 2500 University Drive North West, Calgary, Alberta T2N 1N4, Canada.

**ABSTRACT**

The weakly-bound molecular clusters $(D_2O)_2$-CO and $(D_2O)_3$-CO are observed in the C-O stretch fundamental region ($\approx 2150$ cm$^{-1}$), and their rotationally-resolved infrared spectra yield precise rotational parameters. The corresponding $H_2O$ clusters are also observed, but their bands are broadened by predissociation, preventing detailed analysis. The rotational parameters are insufficient in themselves to determine cluster structures, so *ab initio* calculations are employed, and good agreement between the experimental and theory is found for the most stable cluster isomers, yielding the basic cluster geometries as well as confirming the assignments to $(D_2O)_2$-CO and $(D_2O)_3$-CO. The trimer, $(D_2O)_2$-CO, has a near-planar geometry with one D atom from each $D_2O$ slightly out of the plane. The tetramer, $(D_2O)_3$-CO, has the water molecules arranged in a cyclic quasi-planar ring similar to the water trimer, with the carbon monoxide located 'above' the ring and roughly parallel to its plane. The tunneling effects observed in the free water dimer and trimer are quenched by the presence of CO. The previously observed water-CO dimer together with the trimer and tetramer reported here represent the first three steps in the solvation of carbon monoxide.

# Introduction

High resolution spectroscopy of weakly-bound clusters provides a sensitive and detailed probe of intermolecular interactions (van der Waals and hydrogen bond forces). But detection and analysis of such spectra becomes more challenging as cluster size increases (dimer, trimer, tetramer, …). Quantum chemical calculations can offer important help in the interpretation of experimental spectra, and the analyzed spectrum can then contribute to refinement of the calculations. In the present paper, we report observation of infrared spectra of the trimer $(D_2O)_2$-CO and the tetramer $(D_2O)_3$-CO. *Ab initio* structural calculations, also reported here, played an essential role in understanding the spectra and in establishing the cluster structures responsible for them.

The water-CO dimer is known to have a planar equilibrium structure with approximately collinear heavy atoms (O, C, O) and a hydrogen bond between the water and the carbon of CO. Interchange of the H (or D) atoms gives rise to two resolved tunneling states which correspond to distinct nuclear spin modifications. For the mixed isotope dimer containing HDO, there is no such tunneling but rather two isomers HOD-CO (deuteron bound) and DOH-CO (proton bound). A number of spectroscopic studies of water-CO have been made in the microwave,[1] millimeter wave,[2] and infrared[3-7] regions.

There have been a few high-level ab initio calculations of the water-CO global potential energy surface,[6,8,9] and many of the water-water potential surface.[10] However, in this paper we use new calculations applied directly to mixed water-CO clusters in order to find plausible energy minima to compare with our observed spectra.

## Experimental spectra and analysis

The spectra were recorded in Calgary as described previously[7, 11-13] using a pulsed supersonic slit jet expansion probed by a rapid-scan tunable infrared quantum cascade laser. A typical expansion mixture contained about 0.01% $D_2O$ plus 0.02 – 0.06% CO in helium carrier gas, with a backing pressure of about 10 atmospheres. Wavenumber calibration was carried out by simultaneously recording signals from a fixed etalon and a reference gas cell containing $N_2O$, and spectral simulations were made using the PGOPHER software.[14]

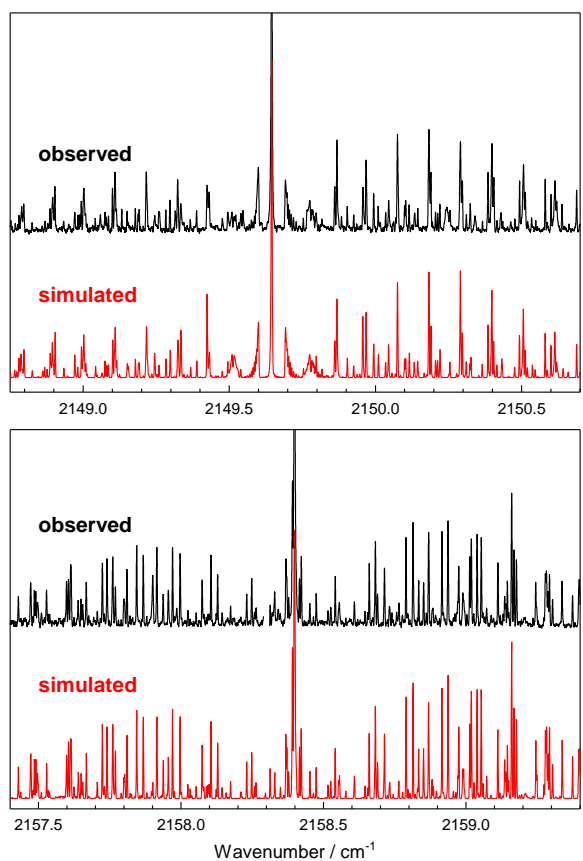

**Figure 1**. Spectra of $(D_2O)_2$-CO (top) and $(D_2O)_3$-CO (bottom) in the C-O stretch region.

While recording spectra[7] of the $D_2O$-CO dimer in the C-O stretch region ($\approx$2150 cm$^{-1}$), we observed two new bands whose appearance and density of lines indicated that they must arise from clusters larger than the dimer. One obvious candidate was the trimer $D_2O$-$(CO)_2$, whose microwave spectrum (in the form of $H_2O$-$(CO)_2$) was studied in 1995 by Peterson et al.[15] However, it was immediately evident that this species was not responsible for either of our new bands because the predictable rotational parameters were not consistent with the observed spectra. By a process of trial, error, and refinement, we eventually managed to obtain excellent fits to both bands in terms of asymmetric rotor models. The key enabler in this process was PGOPHER[14] and its excellent interactive tools. Both spectra are illustrated in Fig. 1, and the parameters resulting from the fits are given in Table 1.

**Table 1.** Experimental and theoretical spectroscopic parameters for $(D_2O)_2$-CO and $(D_2O)_3$-CO. [a]

|  | Experiment | | Theory |
| --- | --- | --- | --- |
|  | Excited state | Ground state | Ground state |
| $(D_2O)_2$-CO | | | |
| $\nu_0$ / cm$^{-1}$ | 2158.3947(1) | | |
| $\Delta\nu_0$ / cm$^{-1}$ | +15.124 | | |
| $A$ / MHz | 5728.20(71) | 5717.98(83) | 5780.2 |
| $B$ / MHz | 2591.90(21) | 2605.57(22) | 2625.3 |
| $C$ / MHz | 1788.52(13) | 1794.08(14) | 1813.8 |
| $D_{JK}$ / kHz | 86(16) | 84(19) | |
| $(D_2O)_3$-CO | | | |
| $\nu_0$ / cm$^{-1}$ | 2149.6460(1) | | |
| $\Delta\nu_0$ / cm$^{-1}$ | +6.375 | | |
| $A$ / MHz | 2938.71(25) | 2942.23(30) | 2907.6 |
| $B$ / MHz | 1622.90(19) | 1625.89(19) | 1667.9 |
| $C$ / MHz | 1595.69(20) | 1590.66(19) | 1642.1 |

[a] Numbers in parentheses are 1σ uncertainties in units of the last quoted digit. $\Delta\nu_0$ is the vibrational shift relative to the free CO molecule. The theoretical values are the equilibrium rotational constants obtained by CCSD(T*)-F12C calculations, augmented by vibrational corrections obtained at DFT level (see text).

The first 'mystery' band is centered at about 2158.4 cm$^{-1}$, which represents a vibrational shift of about +15.1 cm$^{-1}$ relative to the free CO molecule. It lies above the fundamental band of the D$_2$O-CO dimer at 2154.5 cm$^{-1}$. In the analysis of this band, we fitted 144 observed lines in terms of 176 transitions with an rms error of about 0.0003 cm$^{-1}$. The rotational constants from this band, ($A$, $B$, $C$) ≈ (5718, 2606, 1794 MHz), correspond approximately to what one expects for a trimer containing one CO and two D$_2$O molecules. These parameters give a small inertial defect (-0.62 amu Å$^2$), indicating that the structure must be close to planar, at least for all the heavy atoms, C and O. The relative intensity of transitions in the band corresponds to a hybrid structure in which the transition dipole moment in the *a*-inertial direction is roughly twice the magnitude of that in the *b* direction. This gives information about the orientation of CO within the cluster. But even with this constraint, and assuming the formula (D$_2$O)$_2$-CO, the rotational constants were insufficient to establish a unique experimental structure -- there were just too many possibilities.

The second 'mystery' band is centered at about 2149.6 cm$^{-1}$, representing a smaller vibrational shift of about +6.4 cm$^{-1}$. The fit included 125 observed lines assigned to 325 transitions with an rms error of 0.00057 cm$^{-1}$ (there are many blended lines). This is also a hybrid band, with roughly equal *a*- and *b*-dipole contributions, and little or no *c*-dipole contribution. The derived rotational constants, ($A$, $B$, $C$) ≈ (2943, 1625, 1590 MHz), correspond approximately to those expected for the tetramer (D$_2$O)$_3$-CO. There are *many* possible ways to arrange three water molecules and one carbon monoxide, so again we were not able to establish a unique experimental structure without further guidance!

So far, we have presented spectra obtained with D$_2$O. We also observed both new cluster bands using H$_2$O, but unfortunately found that they are significantly broadened, presumably by predissociation in the upper state (that is, shortened lifetimes). This broadening made detailed analysis impossible. Its magnitude for both bands was very approximately 0.009 cm$^{-1}$ (FWHM). For the first band, the origin for the H$_2$O containing species is at 2157.47 cm$^{-1}$, which is about 0.9 cm$^{-1}$ below that of the D$_2$O containing species. For the second band, the origin is 2148.85 cm$^{-1}$, about 0.8 cm$^{-1}$ lower. Interestingly, these H$_2$O to D$_2$O shifts are similar in sign and magnitude to the shift of 0.92 cm$^{-1}$ observed previously[3,7] for the water-CO dimer.

## Computational details

Following a methodology based on previous studies on molecular clusters,[16,17] the potential energy surfaces (PES) of both (D$_2$O)$_2$-CO and (D$_2$O)$_3$-CO clusters were investigated at B2PLYP level of theory[18] in conjunction with the m-aug-cc-pVTZ basis set.[19] To properly take into account dispersion effects,[20,21] the D3BJ corrections[22,23] were used. For both (D$_2$O)$_2$-CO and (D$_2$O)$_3$-CO, 100 different starting geometries were randomly generated and then fully optimized; on each of these structures, subsequent hessian calculations were carried out to confirm that they were true minima. By using the equilibrium rotational constants to cluster these minima, we thus identified three possible low-lying different isomers for (D$_2$O)$_2$-CO and eight for (D$_2$O)$_3$-CO. Through the use of appropriate extrapolation approaches to the complete basis set (CBS) limit, both basis set incompleteness (BSIE) and superposition (BSSE) errors can be accounted for when computing different molecular properties (see, for example, Refs. (24, 25) and references therein).  Therefore, the accurate determination of the binding energies of these clusters at CBS limit were performed employing two different composite schemes. The first one (hereafter labeled as CBS-1) is based on the three-point extrapolation proposed by Feller,[26] the second one, CBS-2, employed the 4-5 inverse polynomial scheme,[27,28] widely used for assessing the energies of water clusters.[29,30] These results are collected in Table 2. The geometry of the most stable structure was further refined by calculations carried out at CCSD(T*)-F12c level of theory.[31] Vibrational corrections to these equilibrium rotational constants were obtained by VPT2 treatment using the anharmonic force constants calculated using the B2PLYP and B3LYP functionals, given their performances reported in the literature for the cubic part of the potential.[32] As done in previous investigation,[17] the potential bias (in the VPT2 step) due to intermolecular motions was taken into account by using a reduced-dimensionality scheme as implemented in an appropriate suite of programs.[33] Further details about the computational methodology and the extrapolation schemes employed in the present work can be found in the supplementary material.

**Table 2.** Binding energies (kcal/mol) extrapolated to the CBS limit for the different $(D_2O)_2$-CO and $(D_2O)_3$-CO structures investigated. [a]

|  | CBS-1 | CBS-2 |
|---|---|---|
| $(D_2O)_2$-CO |  |  |
| 1 | -7.11 (-4.90) | -7.01 (-4.80) |
| 2 | -6.57 (-4.35) | -6.37 (-4.15) |
| 3 | -8.17 (-5.55) | -7.98 (-5.36) |
| $(D_2O)_3$-CO |  |  |
| 1 | -16.75 (-12.00) | -16.47 (-11.72) |
| 2 | -17.96 (-13.19) | -17.77 (-13.00) |
| 3 | -17.99 (-12.90) | -17.75 (-12.66) |
| 4 | -12.77 (-8.74) | -12.54 (-8.51) |
| 5 | -17.93 (-12.98) | -17.68 (-12.73) |
| 6 | -17.11 (-12.35) | -16.93 (-12.17) |
| 7 | -17.97 (-12.90) | -17.73 (-12.66) |
| 8 | -12.45 (-8.59) | -12.24 (-8.38) |

[a] The value computed by including the zero-point vibrational energy are given in parentheses.

## Discussion and conclusions

The calculated rotational parameters for the most stable trimer and tetramer isomers are given in the last column of Table 1. The agreement, though not quite perfect, is definitely sufficient to confirm the assignments to $(D_2O)_2$-CO and $(D_2O)_3$-CO and the basic cluster geometries. Other possible isomers that were investigated not only had higher calculated energies but also had incompatible rotational constants.

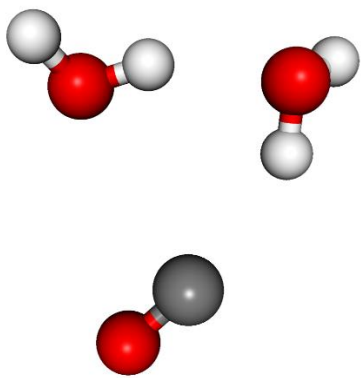

**Figure 2**.  Theoretical structure of the $(D_2O)_2$-CO trimer.

**Table 3.** Calculated structures for $(D_2O)_2$-CO and $(D_2O)_3$-CO. [a]

|   | a | b | c |
|---|---|---|---|
| $(D_2O)_2$-CO | | | |
| C | 1.3735 | -0.5962 | 0.0142 |
| O | 2.3384 | -0.0096 | -0.0121 |
| O | -1.7539 | -1.1394 | -0.1021 |
| D | -0.8104 | -1.3220 | -0.0156 |
| D | -2.1759 | -1.6263 | 0.6098 |
| O | -0.8935 | 1.5854 | 0.0831 |
| D | -1.4237 | 0.7799 | -0.0056 |
| D | -1.3581 | 2.2536 | -0.4247 |
| $(D_2O)_3$-CO | | | |
| C | 1.8538 | 0.6607 | -0.0330 |
| O | 2.5406 | -0.2056 | 0.2049 |
| O | -0.6591 | -1.1418 | -1.2060 |
| D | -0.8359 | -1.3013 | -0.2660 |
| D | -1.0618 | -1.8716 | -1.6780 |
| O | -1.0935 | -0.5340 | 1.4745 |
| D | -1.7608 | -0.6608 | 2.1498 |
| D | -1.2969 | 0.3186 | 1.0532 |
| O | -1.3329 | 1.4897 | -0.4251 |
| D | -0.5040 | 1.9737 | -0.4434 |
| D | -1.1694 | 0.7269 | -1.0002 |

[a] These are principal axis system Cartesian coordinates in units of Angstroms.

The theoretical structures of the most stable trimer and tetramer are listed in Table 3 and shown in Figs. 2 and 3. The trimer geometry has all the heavy atoms, and one D atom from each $D_2O$, in a common plane. This accounts for the small observed inertial defect. One remaining D atom lies above the plane, and the other below it. The CO axis lies at an angle of 31° to the *a*-inertial axis, giving a ratio of 1.6 for the *a*-axis to *b*-axis dipole transition moments, as compared to the observed value of ≈2.0.

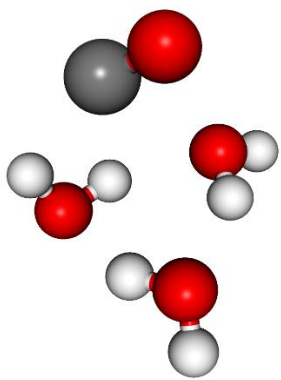

**Figure 3**. Theoretical structure of the $(D_2O)_3$-CO tetramer.

The theoretical tetramer geometry has the three $D_2O$ monomers arranged in a cyclic quasi-planar ring structure, quite similar to the free water trimer.[34] The water O-O distances are 2.79, 2.79, and 2.83 Å, as compared to 2.78 Å for water trimer. The CO monomer is located 'above' the plane defined by the three $D_2O$ O-atoms. The *a*-axis is approximately perpendicular to this plane and passes close to the CO bond. The *b*- and *c*-axes are approximately parallel to this plane, and the *b*-axis is approximately parallel to the CO monomer axis. More precisely, the CO axis makes angles of 53°, 40°, and 78° with the *a*-, *b*-, and *c*-axes. The theoretical ratio of *a* to *b* transition dipoles is 0.79, which agrees well with experiment (this value is used for the simulated spectrum in Fig. 1).

In conclusion, the weakly-bound clusters $(D_2O)_2$-CO and $(D_2O)_3$-CO have been studied experimentally by high-resolution infrared spectroscopy in the C-O stretch fundamental region, and theoretically by means of *ab initio* calculations. The analogous $H_2O$ clusters were also detected, but showed significant ($\approx$0.009 cm$^{-1}$) predissociation broadening which made detailed analysis impossible. The spectra give precise band origins and rotational parameters (*A*, *B*, *C*), but these are not sufficient to specify the cluster structures. However, the calculations give parameters in good agreement with experiment for the lowest energy isomers of each cluster, and thus provide confirmation of the assignments to $(D_2O)_2$-CO and $(D_2O)_3$-CO as well as their essential structural details. The trimer, $(D_2O)_2$-CO, has a near-planar geometry in which one D atom of each $D_2O$ lies slightly out of the plane. The tetramer, $(D_2O)_3$-CO, has the three water molecules in a cyclic quasi-planar ring structure similar to that of the water trimer with the carbon monoxide located 'above' the ring and roughly parallel to its plane. There is no evidence for tunneling effects (such as observed in water dimer and trimer) in the spectra.

**ASSOCIATED CONTENT**

Supporting information includes details on the computational methodology and the geometries determined in the present work.

**ACKNOWLEDGMENTS**

The financial support of the Natural Sciences and Engineering Research Council of Canada is gratefully acknowledged.

# SUPPLEMENTARY MATERIAL

**COMPUTATIONAL METHODOLOGY**

For computing the CBS limit of the binding energies, two different extrapolation schemes were used, both composed of several single point energy calculations carried out on the optimized geometries obtained at B2PLYP-D3BJ level of theory in conjunction with the m-aug-cc-pVTZ basis set.[1] For the composite schemes, the aug-cc-pV$N$Z basis sets with $N$= T, Q and 5 were employed.[2,3] Additional calculations were carried out using the cc-pCVTZ basis set.[4]

The first composite scheme, labelled as CBS-1, is based on separate extrapolation of the CBS limit for the Hartree-Fock (HF-SCF) energy, $E_{CBS}$(HF-SCF) and for the correlation energy computed at MP2 level of theory, $E_{CBS}(corr)$. For $E_{CBS}$(HF-SCF) the expression proposed by Fenner[5] was used, while for $E_{CBS}(corr)$ the following inverse cubic function was used:

$$E_{CBS}(corr) = \frac{N^3 E_N^{corr} - (N-1)^3 E_{N-1}^{corr}}{N^3 - (N-1)^3}.$$

The CBS limit was expressed as:

$$E_{CBS} = E_{CBS}(\text{HF} - \text{SCF}) + E_{CBS}(corr)$$

The effects due to higher-order electron correlation past the MP2 level of theory was taken into account as the difference between the CCSD(T) and MP2 single point energies, computed using the aug-cc-pVTZ basis set.

$$\Delta_{MP2}^{CCSD(T)} = E_{CCSD(T)} - E_{MP2}$$

For computing the core-valence (CV) corrections, two calculations were performed at MP2 level and using the cc-pCVTZ basis set, correlating all the electrons, $E_{ae-MP2}$, and within the frozen-core approximation, $E_{fc-MP2}$.

$$\delta_{MP2}^{CV} = E_{ae-MP2} - E_{fc-MP2}$$

The CBS-1 limit was therefore given as

$$E_{CBS-1} = E_{CBS}(\text{HF} - \text{SCF}) + E_{CBS}(corr) + \Delta_{MP2}^{CCSD(T)} + \delta_{MP2}^{CV}$$

The second composite scheme, labelled as CBS-2, is based on using the 4-5 inverse polynomial extrapolation for the CBS energy at MP2 level,[6,7] according to the following expression:

$$E_{CBS} = E_N(\text{MP2}) + \frac{b}{(N+1)^4} + \frac{c}{(N+1)^5}$$

The corrections due to higher-order electron correlation past the MP2 level of theory, and those due to CV effects, are computed as in CBS-1.

On the basis of these two extrapolation schemes (and taking into account the inclusion of zero-point vibrational correction), the most stable isomers of $(D_2O)_2$-CO and $(D_2O)_3$-CO among the different structures optimized were therefore identified. These two isomers were further optimized at CCSD(T)-F12c level of theory employing the VTZ-F12 and VDZ-F12 basis sets for $(D_2O)_2$-CO and $(D_2O)_3$-CO, respectively. Anharmonic corrections were computed (within the framework of VPT2 theory and employing a reduced dimensionality scheme) using the m-aug-cc-pVTZ basis set and the B2PLYP-D3BJ and B3LYP-D3BJ functionals for $(D_2O)_2$-CO and $(D_2O)_3$-CO, respectively.

All the DFT-D3BJ calculations were performed using the ORCA suite of programs,[8] while MOLPRO was used for CCSD(T*)-F12c computations;[9,10] single point energy calculations were carried out using the Gaussian suite of quantum chemical programs.[11] Geometry optimizations were carried out using the *TightOpt* criteria as implemented in the Orca software.

*****************************************************************************
Optimized geometries (Angstrom) for $(D_2O)_2$-CO at B2PLYP-D3BJ/m-aug-cc-pVTZ level of theory

Geo #1
O   1.622847  -1.147135  -0.100131
D   0.717759  -1.469857  -0.053644
D   2.080711  -1.575062   0.627735
O   0.815533   1.620769   0.080635
D   1.352517   2.247904  -0.407638
D   1.265664   0.769125  -0.015727
C  -2.148802   0.263367  -0.021961
O  -1.503859  -0.667673   0.017127

Geo #2
O   3.184234  -0.652575   0.000170
D   3.712497  -0.401283   0.762621
D   3.712542  -0.401217  -0.762229
O   0.824212   1.093692   0.000100
D   0.002354   0.596186   0.000048
D   1.524999   0.428229   0.000132
C  -2.293791  -0.156094  -0.000137
O  -3.407151  -0.351786  -0.000239

Geo #3
O  -1.753910  -1.139366  -0.102068
D  -0.810431  -1.322000  -0.015634
D  -2.175881  -1.626294   0.609776
O  -0.893528   1.585447   0.083084
D  -1.423694   0.779900  -0.005631
D  -1.358134   2.253610  -0.424746
C   1.373468  -0.596175   0.014192
O   2.338354  -0.009602  -0.012131

***************************************************************************
Optimized geometries (Angstrom) for (D$_2$O)$_3$-CO at B2PLYP-D3BJ/m-aug-cc-pVTZ level of theory

Geo #1
| | | | |
|---|---|---|---|
| C | -1.887305 | 0.548357 | -0.055951 |
| O | -2.853309 | -0.031168 | -0.137439 |
| O | 0.982512 | 1.831276 | 0.069691 |
| D | 1.047221 | 2.396277 | 0.843355 |
| D | 0.079574 | 1.482672 | 0.077903 |
| O | 2.474441 | -0.506361 | -0.120431 |
| D | 2.951774 | -0.532175 | -0.952564 |
| D | 2.085733 | 0.386276 | -0.067218 |
| O | -0.052213 | -1.689620 | 0.153355 |
| D | -0.033812 | -2.533708 | 0.608793 |
| D | 0.881101 | -1.428821 | 0.057338 |

Geo #2
| | | | |
|---|---|---|---|
| C | 1.953123 | 0.547254 | -0.137574 |
| O | 2.621973 | -0.306533 | 0.180859 |
| O | -0.756893 | -1.259900 | -1.089140 |
| D | -1.002931 | -1.288242 | -0.148381 |
| D | -1.189451 | -2.008326 | -1.504978 |
| O | -1.291944 | -0.310001 | 1.460115 |
| D | -2.016230 | -0.283953 | 2.088494 |
| D | -1.353749 | 0.509803 | 0.934175 |
| O | -1.146186 | 1.493853 | -0.642888 |
| D | -0.254439 | 1.853496 | -0.661620 |
| D | -1.070968 | 0.656083 | -1.129436 |

Geo #3
| | | | |
|---|---|---|---|
| C | -2.972141 | -0.410827 | -2.012870 |
| O | -3.453384 | -0.109090 | -2.989400 |
| O | 1.170521 | -1.624118 | 0.887910 |
| D | 1.724245 | -2.264098 | 1.339644 |
| D | 1.112993 | -0.852925 | 1.478764 |
| O | 0.083108 | 0.574168 | 2.225853 |
| D | -0.126512 | 0.762689 | 3.142786 |
| D | -0.714172 | 0.157753 | 1.847437 |
| O | -1.595442 | -1.157729 | 0.872419 |
| D | -1.983832 | -0.980573 | 0.010071 |
| D | -0.745385 | -1.595249 | 0.697386 |

Geo #4
| | | | |
|---|---|---|---|
| C | -1.685950 | -1.054714 | 0.198649 |
| O | -2.699516 | -1.294242 | -0.237052 |
| O | 3.342534 | -0.724293 | -0.450699 |
| D | 3.555350 | -0.252286 | -1.258308 |
| D | 2.548909 | -0.290845 | -0.107375 |
| O | 0.985713 | 0.404753 | 0.844767 |
| D | 0.257260 | -0.232072 | 0.818997 |
| D | 1.237545 | 0.475868 | 1.769829 |
| O | -1.153660 | 1.921707 | -0.336409 |
| D | -0.271706 | 1.729660 | 0.008984 |
| D | -1.053575 | 2.702538 | -0.884927 |

Geo #5
| | | | |
|---|---|---|---|
| C | -2.458665 | -0.953527 | -1.443108 |
| O | -3.307817 | -1.660455 | -1.679196 |
| O | 2.086023 | 1.268387 | 0.992525 |
| D | 2.192642 | 0.589519 | 1.661867 |
| D | 1.161037 | 1.208215 | 0.691205 |
| O | 2.474711 | 1.098466 | -1.761577 |
| D | 2.981849 | 1.804127 | -2.168589 |
| D | 2.654115 | 1.159167 | -0.806296 |
| O | -0.116168 | 1.231845 | -0.690349 |
| D | 0.595136 | 1.186922 | -1.351333 |
| D | -0.762862 | 0.567333 | -0.945150 |

Geo #6
| | | | |
|---|---|---|---|
| C | 0.978824 | -2.036319 | 0.429083 |
| O | 1.993007 | -2.507696 | 0.593977 |
| O | -0.992593 | 0.644713 | 0.736542 |
| D | -0.249102 | 0.575243 | 1.359303 |
| D | -1.065938 | -0.237269 | 0.362312 |
| O | 1.590239 | 0.909214 | 1.825248 |
| D | 1.607869 | 1.598040 | 1.136783 |
| D | 1.918414 | 1.326823 | 2.624575 |
| O | 0.765599 | 2.574956 | -0.231654 |
| D | 0.971327 | 2.654185 | -1.165188 |
| D | -0.017645 | 1.998110 | -0.170982 |

Geo #7
| | | | |
|---|---|---|---|
| C | -2.737332 | 0.430235 | -0.343360 |
| O | -3.707299 | 1.001289 | -0.247173 |
| O | -0.019934 | -1.308447 | -0.941313 |
| D | 0.533731 | -1.399860 | -0.148540 |
| D | -0.786340 | -0.790275 | -0.678257 |
| O | 2.378256 | -0.481702 | -2.036943 |
| D | 1.433562 | -0.698636 | -1.926714 |
| D | 2.674257 | -0.979509 | -2.802052 |
| O | 2.332657 | -1.386371 | 0.600244 |
| D | 2.657703 | -1.045119 | -0.252011 |
| D | 2.740739 | -0.841605 | 1.276120 |

Geo #8
```
C    -0.702320   -0.054020    2.062443
O    -0.411833   -1.141570    2.159385
O    -0.280641    0.518891   -0.832033
D    -0.704813    1.367698   -0.652080
D    -0.432748    0.337857   -1.768591
O    -0.532181   -0.242916   -3.650429
D     0.346473   -0.440299   -3.985662
D    -1.038817   -1.048566   -3.782531
O    -1.469372    2.688763    0.666560
D    -0.995997    3.480190    0.933677
D    -1.277752    2.033972    1.349261
```